\documentclass[11pt,a4paper]{article}
\usepackage{jheppub}
\usepackage{natbib}
\allowdisplaybreaks

\newcommand{\nablaslash}{\nabla{\!\!\!\!\slash}}

\newcommand{\pslash}{p{\hspace{-5pt}\slash}}
\newcommand{\qslash}{q{\hspace{-5pt}\slash}}
\newcommand{\bare}{\textrm{B}}

\begin{document}

\title{Effective action for the Yukawa model in curved spacetime}

\author{David J. Toms}
\emailAdd{david.toms@newcastle.ac.uk}
\affiliation{
School of Mathematics, Statistics and Physics,
Newcastle University,
Newcastle upon Tyne, U.K. NE1 7RU}

\date{\today}

\abstract
{We consider the one-loop renormalization of a real scalar field interacting with a Dirac spinor field in curved spacetime. A general Yukawa interaction is considered which includes both a scalar and a pseudoscalar coupling. The scalar field is assumed to be non-minimally coupled to the gravitational field and to have a general quartic self-interaction potential. All of the one-loop renormalization group functions are evaluated and in the special case where there is no mass scale present in the classical theory (apart from the fields) we evaluate the one-loop effective action up to and including order $R^2$ in the curvature. In the case where the fermion is massive we include a pseudoscalar mass term in $\gamma_5$ and we show that although the $\gamma_5$ term can be removed by a redefinition of the spinor field an anomaly in the effective action arises that is related to the familiar axial current anomaly.}


\maketitle

\section{Introduction}\label{sec-intro}

The general area of quantum field theory in curved spacetime is now well established. (Several reviews emphasizing various aspects are \cite{Gibbons,BirrellandDavies,fulling1989aspects,Buchbinderbook, Ford,ParkerTomsbook} and earlier influential treatments include \cite{DeWittdynamical,dewitt1975quantum}.) One model that has not been analyzed in the same depth as say the interacting scalar field or Yang-Mills theory in curved spacetime is that of a scalar field interacting with a Dirac spinor through a Yukawa interaction. There is some previous work that we will mention briefly.

The renormalization group behaviour of various grand unified models in curved spacetime, which incorporate Yukawa interactions, was considered in \cite{parker1984renormalization}. It was first noted here that the non-minimal coupling between the scalar field and the curvature (that we will generically refer to as $\xi R\varphi^2$) only involves the non-minimal parameter in the combination $(\xi-1/6)$ in the associated renormalization group function.

A number of papers \cite{shapiro1989asymptotic,odintsov1992general,elizalde1994renormalization,elizalde1995higgs, elizalde1995improved,elizalde1995renormalization,Shapiro2011PLB} study special cases of the theory that we will consider here. Some of these special cases include massless scalars or spinors, constant background curvature, or constant background scalar fields. The Yukawa interaction is chosen to be of the scalar, as opposed to the pseudoscalar (involving $\gamma_5$), type. (See \eqref{2.2b} below for details.) In the case of massless fermions this choice makes no difference as we will discuss in Sec.~\ref{secanomaly}. Effective potentials that include terms linear in the curvature are calculated in some cases as are renormalization group improved effective potentials. Some work includes contributions from $R^2$ gravity as well. More recent work \cite{herranen2014spacetime,Cz,markkanen20181} has studied the role of the effective potential in a sector of the standard model.

There is also a body or work \cite{ProkopecWoodard,Garbrechtfermion,Garbrecht,Miao} on calculations in de Sitter spacetime that include a Yukawa interaction between a scalar and a spinor. A general conformally flat background has also been explored~\cite{DuffyWoodard}. One of the important consequences of this work is the generation of a mass term for a fermion field that has no bare mass. The Yukawa interaction is responsible for this.

Gravitational corrections to the Yukawa coupling has also been a very active area of research recently. This has been done within the asymptotic safety program \cite{Zanusso,Eichhornetal1,Eichhornetal3,oda2016non,Eichhornetal2} and also using more standard perturbative methods~\cite{RodigastSchuster,narain2017exorcising,Martins2017PLB,Martins2018JCAP,Martins2018EPJC}. (See also the illuminating discussion of \cite{Anberetal1}.)

The purpose of the present paper is to present a complete analysis of the one-loop effective action for a real scalar field interacting with a Dirac spinor in a general curved spacetime. We will allow both fields to be massive with the scalar field non-minimally coupled to the scalar curvature. We consider a general quartic potential for the scalar field and do not impose the $\varphi\rightarrow -\varphi$ symmetry since this is broken by the Yukawa term if the fermions are massive. We will allow both a scalar and pseudoscalar Yukawa interaction and also consider both scalar and pseudoscalar mass terms for the spinor (see \eqref{2.2b} below) for reasons that will be discussed later. We do not assume that the background scalar field is constant or that the spacetime has a constant curvature. The complete one-loop renormalization of the theory is considered and we calculate all of the renormalization group functions. In the special case where the classical theory has no mass scales other than those in the fields we calculate terms in the effective action up to order $R^2$. Field redefinitions that can transform away the pseudoscalar mass term (or alternatively the pseudoscalar Yukawa coupling) are considered in Sec.~\ref{secanomaly} and a resulting anomaly in the effective action is calculated. An appendix gives a very brief discussion of the renormalization group solutions for the Yukawa and quartic scalar field coupling constants.

\section{The Yukawa model in curved spacetime}\label{sec2}

We will consider a real scalar field $\Phi(x)$ coupled to a Dirac spinor $\Psi(x)$ and allow the possibility of both a scalar and a pseudoscalar Yukawa coupling. The spacetime dimension will be exclusively four and we will adopt dimensional regularization~\cite{tHooftandVeltman}. All of our conventions for curvature and spinors will follow those in Parker and Toms~\cite{ParkerTomsbook}. We will use the vierbein formalism to relate the Dirac matrices to a local orthonormal frame with a Minkowski metric so that the usual Dirac algebra can be applied as in, for example, Bjorken and Drell~\cite{BjorkenandDrell}. The bare action will be taken to be (with the subscript `B' signifying a bare quantity)
\begin{equation}
S=S_{\textrm {scalar}}+S_{\textrm{spinor}}+S_{\textrm{grav}},\label{2.1}
\end{equation}
where
\begin{subequations}\label{2.2}
\begin{align}
S_{\textrm{scalar}}&=\int dv_x \Big\lbrack \frac{1}{2}\partial^\mu\Phi_{\textrm{B}} \partial_\mu\Phi_{\textrm{B}} -\frac{1}{2}m_{\textrm{s\,B}}^2\, \Phi_{\textrm{B}}^2 -\frac{1}{2}\xi_{\textrm{B}}\,R \,\Phi_{\textrm{B}}^2 -\frac{\lambda_{\textrm{B}}}{4!}\,\Phi_{\textrm{B}}^4\nonumber\\
&\quad -\frac{1}{3!}\eta_{\textrm{B}}\,\Phi_{\textrm{B}}^3-\tau_{\textrm{B}}\,\Phi_{\textrm{B}}-\gamma_{\textrm{B}}\,R\,\Phi_{\textrm{B}} \Big\rbrack,\label{2.2a}\\
S_{\textrm{spinor}}&=\int dv_x\Big\lbrack \bar{\Psi}_{\textrm{B}}(i\gamma^\mu\nabla_\mu-m_{0\,\textrm{B}}-im_{5\,\textrm{B}}\gamma_5)\Psi_{\textrm{B}}-\Phi_{\textrm{B}}\bar{\Psi}_{\textrm{B}}(w_{\textrm{B}}+iw_{5\,\textrm{B}}\gamma_5)\Psi_{\textrm{B}}\big\rbrack,\label{2.2b}\\
S_{\textrm{grav}}&=\int dv_x\Big(\Lambda_{\textrm{B}}+\kappa_{\textrm{B}}\,R+\alpha_{1\,\textrm{B}}\,R^{\mu\nu\lambda\sigma}R_{\mu\nu\lambda\sigma} + \alpha_{2\,\textrm{B}}\,R^{\mu\nu}R_{\mu\nu} + \alpha_{3\,\textrm{B}}\,R^2 \Big).\label{2.2c}
\end{align}
\end{subequations}
We use $dv_x$ to stand for the invariant spacetime volume element: $dv_x=|\det g_{\mu\nu}(x)|^{1/2}d^nx$. 

The first line of $S_{\textrm{scalar}}$ is the usual one for a real scalar field with a general non-minimal coupling to the spacetime curvature. The terms in the second line of $S_{\textrm{scalar}}$ are necessary for renormalizability in the case where the  field does not have the discrete $\mathbb{Z}_2$ symmetry $\Phi\rightarrow-\Phi$ as is the case here in general. This will be seen later in Secs.~\ref{sec3}--\ref{sec5}. (We use the nomenclature of \cite{ParkerTomsbook}.) In $S_{\textrm{spinor}}$ we include the usual mass term for the Dirac field, $m_0$, as well as a pseudoscalar mass term $m_5$. It is possible to remove the term in $m_5$ by a chiral transformation of the spinor field; however we will not do this here. This will be discussed in more detail in Sec.~\ref{secanomaly} where it will be shown to lead to an anomaly in the quantum theory. This anomaly does not affect the divergent part of the effective action or counterterms and by keeping the action in the form of \eqref{2.2c} we will show this explicitly. The covariant derivative $\nabla_\mu$ includes the usual spin connection for the Dirac field. The second term in $S_{\textrm{spinor}}$ is the Yukawa coupling of the Dirac spinor to the scalar field and we allow for both a scalar coupling with coupling constant $w$ as well as a pseudoscalar coupling with coupling constant $w_5$. Here $\gamma^\mu$ are the spacetime dependent Dirac matrices~\cite{ParkerTomsbook} defined in terms of the usual Minkowski ones $\gamma^a$ by $\gamma^\mu=e_{a}{}^{\mu} \gamma^a$ with $e_{a}{}^{\mu}$ the vierbein. (Latin letters will denote orthonormal frame indices.) $\gamma_5$ is the usual expression~\cite{BjorkenandDrell} which is Hermitian, obeys $\gamma_5^2=I$ with $I$ the identity matrix, and constant (since it is defined in terms of the local orthonormal frame $\gamma$-matrices). The factors of $i$ in \eqref{2.2b} ensure that the action is real. The gravitational part of the action, $S_{\textrm{grav}}$, is required to deal with the vacuum part of the effective action in the usual way~\cite{BirrellandDavies,ParkerTomsbook}.

We will adopt DeWitt's~\cite{DeWittdynamical} background field method and compute the effective action. We are primarily interested in the renormalization so we will be concerned just with the pole parts of the effective action and we will only work at one-loop order. The full discussion of the renormalization counterterms and the relationship between the bare and renormalized quantities in the theory will be deferred until Sec.~\ref{sec5}. To one-loop order to compute the one-loop part of the effective action we can take all expressions to be renormalized ones and these are indicated without the `B' subscript appended. We will initially concentrate just on the scalar sector of the theory by allowing only the background scalar field to be non-zero. The case of a background Dirac field will be deferred until Sec.~\ref{sec4.2}. If we let $\varphi(x)$ be the background scalar field then a functional evaluation of the effective action results in the formal expression
\begin{align}
\Gamma_{\textrm{1-loop}}&=\frac{i}{2}\,\ln\det\left(\Box+m_{\textrm{s}}^2+\xi R+\frac{\lambda}{2}\varphi^2+\eta\varphi \right)\nonumber\\
&\quad -i\ln\det\left(i\nablaslash - M_0-iM_5\gamma_5\right),\label{2.3}
\end{align}
where we have defined
\begin{subequations}\label{2.4}
\begin{align}
M_0&=m_0+w\varphi,\label{2.4a}\\
M_5&=m_5+w_5\varphi.\label{2.4b}
\end{align}
\end{subequations}
We do not treat the background field as constant here. This is important if we are to calculate any field renormalization factors. The first term in \eqref{2.3} arises from the functional integral over the scalar field and the second one from the integration over the Dirac field using the usual definitions for Grassmann variables. The usual $i\epsilon$ prescription to ensure Feynman boundary conditions is understood.

We will present two different methods for evaluating the pole part of \eqref{2.3}. The simplest and most elegant method consists of using the heat kernel method~\cite{DeWittdynamical,fulling1989aspects,avramidi2000heat,vassilevich2003heat,ParkerTomsbook,kirsten2010spectral} and we will describe this in the next section. As an alternative approach, and as a check on the heat kernel calculation, we will consider the use of the local momentum space method pioneered by Bunch and Parker~\cite{BunchParker} in Sec.~\ref{sec4}. This method is closer to the types of calculation more familiar in flat spacetime quantum field theory.

\section{Heat kernel method}\label{sec3}

We will use the heat kernel method in this section along with dimensional regularization to compute the pole part of the one-loop effective action. For any operator of the form $D^2+Q$, where $D_\mu$ represents any type of covariant derivative and $Q$ is matrix-valued but not necessarily constant, we have~\cite{ParkerTomsbook}
\begin{equation}
{\textrm{PP}}\Big\lbrace \frac{i}{2}\,\ln\det(D^2+Q) \Big\rbrace=-\frac{1}{16\pi^2\epsilon}\int dv_x\,{\textrm{tr}}(E_2),\label{3.1}
\end{equation}
where we use ${\textrm{PP}}\lbrace \cdots\rbrace$ to denote the pole part of any expression and $E_2$ is a heat kernel coefficient. If we ignore terms that are total derivatives,
\begin{align}
E_2&=\left(\frac{1}{72}\,R^2-\frac{1}{180}\,R^{\mu\nu}R_{\mu\nu}+\frac{1}{180}\,R^{\mu\nu\lambda\sigma}R_{\mu\nu\lambda\sigma}\right)\,I\nonumber\\
&\quad +\frac{1}{12}\,W^{\mu\nu}W_{\mu\nu}+\frac{1}{2}\,Q^2-\frac{1}{6}\,R\,Q,\label{3.2}
\end{align}
where $W_{\mu\nu}$ is the curvature of the covariant derivative $D_\mu$ defined by
\begin{equation}
W_{\mu\nu}=\lbrack D_\mu,D_\nu\rbrack.\label{3.3}
\end{equation}
The most general derivation of $E_2$ was given by Gilkey~\cite{Gilkey75,Gilkey79} but earlier derivations in some cases were given by DeWitt~\cite{DeWittdynamical}. References~\cite{fulling1989aspects,avramidi2000heat,vassilevich2003heat,ParkerTomsbook,kirsten2010spectral} contain more details and history.

The first term of \eqref{2.3} that represents the functional integral over the scalar field is easily dealt with by first noting that for scalar fields $D_\mu=\nabla_\mu$ and $W_{\mu\nu}=0$ and second that we can read off $Q$ to be
\begin{equation}
Q=m_{\textrm{s}}^2+\xi R+\frac{\lambda}{2}\varphi^2+\eta\varphi.\label{3.4}
\end{equation}
We then find that 
\begin{align}
{\textrm{PP}}\left\lbrace \frac{i}{2}\,\ln\det\left(\Box+m_{\textrm{s}}^2+\xi R+\frac{\lambda}{2}\varphi^2+\eta\varphi \right)\right\rbrace&=-\frac{1}{16\pi^2\epsilon}\int dv_x\,\Big\lbrack\frac{1}{2}(\xi-\frac{1}{6})^2R^2  \nonumber\\
&\hspace{-72pt}- \frac{1}{180}R^{\mu\nu}R_{\mu\nu}+ \frac{1}{180}R^{\mu\nu\lambda\sigma}R_{\mu\nu\lambda\sigma}+\frac{1}{2}m_{\textrm{s}}^4 + (\xi-\frac{1}{6})m_{\textrm{s}}^2 R \nonumber\\
&\hspace{-72pt} + \eta(\xi-\frac{1}{6})R\varphi +\frac{1}{2}(\lambda m_{\textrm{s}}^2+\eta^2)\varphi^2 + \frac{\lambda}{2}(\xi-\frac{1}{6})R\varphi^2\nonumber\\
&\hspace{-72pt}+ \eta m_{\textrm{s}}^2 \varphi+\frac{1}{2}\lambda\eta\varphi^3+\frac{1}{8}\lambda^2\varphi^4\Big\rbrack.\label{3.5}
\end{align}
It is noteworthy that the dependence on the scalar curvature $R$ involves $\xi$ only in the combination $(\xi-1/6)$ as proposed in \cite{ParkerToms85a}. This was later proven to all orders in the heat kernel expansion~\cite{JackParker85}. The case where $\eta=0$ was first given by the method we have just described in \cite{TomsPRDscalar}. In the $\eta=0$ case, meaning that the cubic interaction is absent from the original action \eqref{2.2a}, it can be seen that the pole terms that are odd in $\varphi$ (that comprise $\varphi$, $R\varphi$ and $\varphi^3$) all vanish and the result is symmetric under the ${\mathbb{Z}}_2$ reflection symmetry.  The multicomponent scalar field version can be found in \cite{ParkerTomsbook}.

The second term of \eqref{2.3} is not of the form \eqref{3.1} since the Dirac operator is first order. However it can be manipulated into a second order form where we will be able to use \eqref{3.1} and \eqref{3.2}. One way to do this is to introduce a factor of $\gamma_5^2=I$ in front of the Dirac operator, and move one of the $\gamma_5$'s through the Dirac operator using $\gamma_5\nablaslash=-\nablaslash \gamma_5$. It then follows that
\begin{equation}
\ln\det\left(i\nablaslash - M_0-iM_5\gamma_5\right)=\ln\det\left(-i\nablaslash - M_0-iM_5\gamma_5\right).\label{3.6}
\end{equation} 
An equivalent way to deduce this is to note that in any spacetime of even dimension the Dirac matrices $\gamma_\mu$ and their negatives $-\gamma_\mu$ give equivalent representations of the Dirac algebra $\gamma_\mu\gamma_\nu+\gamma_\nu\gamma_\mu=2g_{\mu\nu}I$. Thus for even spacetime dimensions no results should depend on the sign of the Dirac matrices. (See \cite{ParkerTomsbook} for a pedagogical treatment.) A third way is to introduce the Dirac propagator and to define it in terms of a new Green's function that obeys a second order equation as done by DeWitt~\cite{DeWittdynamical} for example. Regardless of the reasoning that leads to \eqref{3.6} we can conclude that
\begin{equation}
\ln\det\left(i\nablaslash - M_0-iM_5\gamma_5\right)=\frac{1}{2}\ln\det\left\lbrack\left(-i\nablaslash - M_0-iM_5\gamma_5\right)\left(i\nablaslash - M_0-iM_5\gamma_5\right)\right\rbrack.\label{3.7}
\end{equation}
The operator appearing on the right hand side of \eqref{3.7} is now second order and after a bit more work we can apply the results of \eqref{3.1} and \eqref{3.2}.

If we expand out the product of operators on the right hand side of \eqref{3.7} being careful to note that neither $M_0$ nor $M_5$ defined in \eqref{2.4} are to be regarded as constant, and using $\nablaslash^2=\Box+\frac{1}{4}RI$, we find
\begin{align}
\left(-i\nablaslash - M_0-iM_5\gamma_5\right)\left(i\nablaslash - M_0-iM_5\gamma_5\right) &= \Box-2M_5\gamma_5\nablaslash-i\gamma^\mu\partial_\mu M_0+\gamma^\mu\gamma_5\partial_\mu M_5\nonumber\\
&\quad+M_0^2I-M_5^2I+2iM_0M_5\gamma_5+\frac{1}{4}RI.\label{3.8}
\end{align}
Because of the presence of the term in $\nablaslash$ on the right hand side this is still not of the form $D^2+Q$  that we require; however it can be put into the required form by defining a new covariant derivative by
\begin{equation}
D_\mu=\nabla_\mu-M_5\gamma_5\gamma_\mu.\label{3.9}
\end{equation}
With this choice we find
\begin{equation}
\left(-i\nablaslash - M_0-iM_5\gamma_5\right)\left(i\nablaslash - M_0-iM_5\gamma_5\right) = D^2+Q,\label{3.10}
\end{equation}
as required where
\begin{equation}
Q=\left( M_0^2+3M_5^2+\frac{1}{4}R\right)I+2iM_0M_5\gamma_5-i\gamma^\mu\partial_\mu M_0.\label{3.11}
\end{equation}
Using \eqref{3.9} it can be shown that the curvature of the covariant derivative defined in \eqref{3.3} is given by
\begin{equation}
W_{\mu\nu}=-\frac{1}{4}R_{\mu\nu\lambda\sigma}\gamma^\lambda\gamma^\sigma-\partial_\mu M_5\gamma_5\gamma_\nu + \partial_\nu M_5\gamma_5\gamma_\mu-M_5^2\lbrack\gamma_\mu,\gamma_\nu\rbrack.\label{3.12}
\end{equation}
It is now just a matter of algebra to show that
\begin{align*}
{\textrm{PP}}\Big\lbrace i\ln\det\left(i\nablaslash - M_0-iM_5\gamma_5\right)\Big\rbrace &= -\frac{1}{16\pi^2\epsilon}\int dv_x\,\Big(\frac{1}{72}R^2-\frac{1}{45}R_{\mu\nu}R^{\mu\nu}- \frac{7}{360}R_{\mu\nu\lambda\sigma}R^{\mu\nu\lambda\sigma}\\
&\hspace{-48pt}+2(M_0^2+M_5^2)^2+\frac{1}{3}(M_0^2+M_5^2)R-2\partial^\mu M_0\partial_\mu M_0 -2\partial^\mu M_5\partial_\mu M_5\Big).
\end{align*}
This result holds for any values of $M_0$ and $M_5$. If we now substitute in the particular values given in \eqref{2.4} we find
\begin{align}
{\textrm{PP}}\Big\lbrace - i\ln\det\left(i\nablaslash - M_0-iM_5\gamma_5\right)\Big\rbrace
&=\frac{1}{16\pi^2\epsilon}\int dv_x\,\Big\lbrace\frac{1}{72}R^2-\frac{1}{45}R_{\mu\nu}R^{\mu\nu}\nonumber\\
&\hspace{-48pt}- \frac{7}{360}R_{\mu\nu\lambda\sigma}R^{\mu\nu\lambda\sigma}+\frac{1}{3}(m_0^2+m_5^2)\,R+\frac{2}{3}(wm_0+w_5m_5 )\, R\,\varphi \nonumber\\
&\hspace{-48pt}+ \frac{1}{3}(w^2+w_5^2)\, R\, \varphi^2-2\,(w^2+w_5^2)\,\partial^\mu\varphi\partial_\mu\varphi+2\, ({m_0}^2+{m_5}^2)^2\nonumber\\
&\hspace{-48pt} + 8\,({m_0}^2+{m_5}^2) ({m_0} w+{m_5} {w_5})\,\varphi\nonumber\\
&\hspace{-48pt} + 4\, \left\lbrack{m_0}^2 (3 w^2+{w_5}^2)+4 {m_0} {m_5} w {w_5}+{m_5}^2 (w^2+3 {w_5}^2)\right\rbrack\, \varphi^2\nonumber\\
&\hspace{-48pt} + 8\, (wm_0+w_5m_5) (w^2 + w_5^2)\, \varphi^3 + 2\,(w^2+{w_5}^2)^2\,\varphi^4\Big\rbrace .\label{3.14}
\end{align}

Combining \eqref{3.5} and \eqref{3.14} will give us the pole part of the one-loop part of the effective action defined by \eqref{2.3}. Even if we take $\eta=0$ so that the scalar field part of the action in \eqref{2.2a} is ${\mathbb{Z}}_2$ invariant this invariance will be broken by the Yukawa term in general and pole terms odd in $\varphi$ will result in general. The exception to this is when we take $w=0$ and $m_5=0$ in the Dirac action \eqref{2.2b}. In this case the theory is symmetric under the parity transformation~\cite{peskin1995introduction} $\Psi(t,{\mathbf x})\rightarrow\gamma^0\Psi(t,-{\mathbf x})$ and $\Phi(t,{\mathbf x})\rightarrow -\Phi(t,-{\mathbf x})$.

\section{Local momentum space method}\label{sec4}

In this section we will derive the pole part of the one-loop effective action that involves the interacting terms of the Lagrangian using a different method that is more closely related to the use of traditional Feynman diagrams. (Of course the usual perturbative expansion using Feynman diagrams could also be used.) The vacuum part of the pole part of the effective action, the terms in \eqref{3.5} and \eqref{3.14} that do not depend on the background scalar field $\varphi$, are the same as those for the free theory so are well known. (See \cite{BirrellandDavies} or \cite{ParkerTomsbook} and references therein for example.) Similarly, the scalar field results in \eqref{3.14} agree with \cite{ParkerTomsbook}. For this reason we will concentrate on terms that depend only on the background scalar field that arise from the Dirac part of the effective action found in the second term of \eqref{2.3} in Sec.~\ref{sec4.1} and for terms that depend on the background Dirac field in Sec.~\ref{sec4.2}. It would be possible to deal with both the scalar field and the spinor field as a multicomponent field with an associated multicomponent Green function in a similar manner to the gravity-QED calculation of \cite{toms2010quantum,Tomsquadratic} but we will not do this here for simplicity. 

The most effective way to deal with the curved spacetime aspect of the calculation is to use the local momentum space approach initiated by Bunch and Parker~\cite{BunchParker}. This has the advantage of looking like the calculation that would be done in flat Minkowski spacetime with the exception that the propagators, or Green's functions, are more complicated. A review and applications of the method may be found in \cite{ParkerTomsbook}. Another advantage is that we can deal directly with the usual first order form for the Dirac operator without the necessity of the manipulations that lead to \eqref{3.10}.

\subsection{Scalar sector}\label{sec4.1}

We begin with the second term of \eqref{2.3},
\begin{equation}
\Gamma_{\textrm{D}}=-i\ln\det\left(i\nablaslash - m_0-im_5\gamma_5-w\varphi-iw_5\varphi\gamma_5\right).\label{4.1}
\end{equation}
We have used $M_0$ and $M_5$ as given in \eqref{2.4}. For brevity let
\begin{equation}
m_{\psi}=m_0+im_5\gamma_5.\label{4.2}
\end{equation}
Define the Dirac field Green's function $S(x,x^\prime)$ by
\begin{equation}
(i\nablaslash - m_\psi)S(x,x^\prime)=-I\delta(x,x^\prime).\label{4.3}
\end{equation}
A simple manipulation of \eqref{4.1} shows that
\begin{equation}
\Gamma_{\textrm{D}}=-i\ln\det(i\nablaslash - m_\psi)-i\,{\textrm{Tr}}\ln(I+X),\label{4.4}
\end{equation}
where
\begin{equation}
X(x,x^\prime)=S(x,x^\prime)(w+iw_5\gamma_5)\varphi(x^\prime).\label{4.5}
\end{equation}
We use ${\textrm{Tr}}$ to denote the functional trace of the enclosed expression and reserve ${\textrm{tr}}$ for the Dirac trace. For example, ${\textrm{Tr}} X=\int dv_x\, {\textrm{tr}}X(x,x)$.

All of the dependence on the background scalar field $\varphi$ is contained in $X$ and is solely in the second term of \eqref{4.4}. The first term of \eqref{4.4} contains just the vacuum part of \eqref{3.14} and we will not evaluate this again. We can expand the second term of \eqref{4.4} in powers of $X$:
\begin{align}
\Gamma_{\textrm{D}\,2}&=-i\,{\textrm{Tr}}\ln(I+X)\nonumber\\
&=i\sum_{n=1}^{\infty}\frac{(-1)^n}{n}{\textrm{Tr}}(X^n).\label{4.6}
\end{align}
Because we are just concerned with the pole terms only the first four terms in the sum need to be considered. This can be seen by simple power counting because we are only working at one-loop order. The leading part of the Dirac Green's function will be that found in flat spacetime. (As we will show below the curvature corrections to the flat spacetime contribution are more convergent.) The momentum space dependence of $S(x,x^\prime)$ will be inverse momentum just as in flat spacetime so once we have an expression with more than four factors of $S(x,x^\prime)$ the momentum space integrations will not contain any poles. It therefore follows that 
\begin{equation}
{\textrm{PP}}\Big\lbrace \Gamma_{\textrm{D}\,2}\Big\rbrace = {\textrm{PP}}\Big\lbrace \Gamma_{\textrm{D}\,2}^{(1)}+ \Gamma_{\textrm{D}\,2}^{(2)}+\Gamma_{\textrm{D}\,2}^{(3)}+\Gamma_{\textrm{D}\,2}^{(4)}\Big\rbrace,\label{4.7}
\end{equation}
where
\begin{subequations}\label{4.8}
\begin{align}
\Gamma_{\textrm{D}\,2}^{(1)}&=-i\,{\textrm{Tr}}(X),\label{4.8a}\\
\Gamma_{\textrm{D}\,2}^{(2)}&=\frac{i}{2}\,{\textrm{Tr}}(X^2),\label{4.8b}\\
\Gamma_{\textrm{D}\,2}^{(3)}&=-\frac{i}{3}\,{\textrm{Tr}}(X^3),\label{4.8c}\\
\Gamma_{\textrm{D}\,2}^{(4)}&=\frac{i}{4}\,{\textrm{Tr}}(X^4).\label{4.8d}
\end{align}
\end{subequations}

We will now obtain the leading terms in the local momentum space expansion for the Dirac Green's function defined in \eqref{4.3}. The case where $m_5=0$ was done in the original calculation by Bunch and Parker~\citep{BunchParker}. 

Begin by choosing the origin of Riemann normal coordinates to be $x^{\prime\mu}$ and set $x^\mu=x^{\prime\mu}+y^\mu$. The differential operator $i\nablaslash-m_\psi$ that defines the Green's function in \eqref{4.3} is then expanded in powers of $y^\mu$. We need to use~\cite{ParkerTomsbook} $\nabla_\mu=\partial_\mu+\frac{1}{4}\omega_{\mu\,ab}\gamma^a\gamma^b$ where 
\begin{equation}
\omega_{\mu}{}^{a}{}_{b}=-e_{b}{}^{\nu}\partial_\mu e^{a}{}_{\nu}+\Gamma^{\lambda}_{\mu\nu}e_{b}{}^{\nu}e^{a}{}_{\lambda}\label{4.9}
\end{equation}
is the spin connection. The first terms in the Riemann normal coordinate expansion of the metric, vierbein, and Christoffel connection are
\begin{align}
g_{\mu\nu}(x)&=\eta_{\mu\nu}+\frac{1}{3}R_{\mu\alpha\nu\beta}y^\alpha y^\beta+\cdots,\label{4.10}\\
e^{a}{}_{\mu}(x)&=e^{a}{}_{\lambda}(x^\prime)\left( \delta^{\lambda}_{\mu}+\frac{1}{6} R^{\lambda}{}_{\alpha\mu\beta}y^\alpha y^\beta+\cdots\right),\label{4.11}\\
\Gamma^{\lambda}_{\mu\nu}(x)&=\frac{1}{3}(R^{\lambda}{}_{\mu\nu\alpha}+R^{\lambda}{}_{\nu\mu\alpha})y^\alpha+\cdots.\label{4.12}
\end{align}
The Riemann tensor appearing in \eqref{4.10}--\eqref{4.12} is evaluated at the origin of Riemann normal coordinates. When these expansions are used in \eqref{4.9} we find that the spin connection has the expansion
\begin{equation}
\omega_{\mu}{}^{a}{}_{b}(x)=\frac{1}{2}e^{a}{}_{\lambda}(x^\prime)e_{b}{}^{\nu}(x^\prime)R^{\lambda}{}_{\nu\mu\alpha}y^\alpha+\cdots.\label{4.13}
\end{equation}
For the spacetime dependent Dirac matrices we use $\gamma_\mu(x)=e^{a}{}_{\mu}(x)\gamma_a$ along with \eqref{4.10} and \eqref{4.11} to find
\begin{equation}
\gamma^{\mu}(x)=\gamma^{\prime\lambda}\left(\delta^{\mu}_{\lambda}-\frac{1}{6}R^{\mu}{}_{\alpha\lambda\beta}y^\alpha y^\beta+\cdots\right),\label{4.14}
\end{equation}
where we shorten $\gamma^\mu(x^\prime)$ to just $\gamma^{\prime\mu}$. Because the metric tensor at $x^\prime$ is just the Minkowski metric, $\gamma^{\prime\mu}$ are just the usual flat spacetime Dirac matrices. A short calculation using the expansions just described results in
\begin{equation}
\nablaslash=\gamma^{\prime\mu}\partial_\mu-\frac{1}{6}R^{\mu}{}_{\alpha\lambda\beta}y^\alpha y^\beta \gamma^{\prime\lambda}\partial_\mu + \frac{1}{4}R_{\mu\nu}y^\mu\gamma^{\prime\nu}+\cdots.\label{4.15}
\end{equation}

We now expand the Dirac field Green's function about the origin of Riemann normal coordinates as
\begin{equation}
S(x,x^\prime)=\int\frac{d^np}{(2\pi)^n}\,e^{ip_\mu y^\mu}\,\left\lbrack S_0(p) + S_1(p;x^\prime) + \cdots\right\rbrack,\label{4.16}
\end{equation}
where $S_0$ is the flat spacetime expression, with $S_1$ and higher order terms corrections to the flat spacetime result due to spacetime curvature. We will choose $S_1$ to be linear in the curvature. The next terms would involve derivatives of the curvature and higher order curvature expressions that we will not need here. After some calculation using \eqref{4.3} along with the expansions \eqref{4.15} and \eqref{4.16} it can be shown that ($\pslash=\gamma^{\prime\mu}p_\mu$ here)
\begin{align}
S_0(p)&=(\pslash+m_\psi)^{-1}\nonumber\\
&=\frac{\pslash-\widetilde{m}_\psi}{p^2-m_0^2-m_5^2},\label{4.17}
\end{align}
where 
\begin{equation}
\widetilde{m}_\psi=m_0-im_5\gamma_5,\label{4.18}
\end{equation}
and that
\begin{align}
S_1(p;x^\prime)&=-\frac{1}{6}R^{\mu}{}_{\alpha\lambda\beta}S_0(p)\gamma^{\prime\lambda} \frac{\partial^2}{\partial p_\alpha\partial p_\beta}\lbrack p_\mu S_0(p)\rbrack - \frac{1}{4}R_{\alpha\nu}S_0(p)\gamma^{\prime\nu} \frac{\partial}{\partial p_\alpha}S_0(p)\nonumber\\
&= \frac{1}{3}R_{\mu\alpha\lambda\beta}p^\mu p^\beta (\pslash-\widetilde{m}_\psi) \gamma^{\prime\lambda}\gamma^{\prime\alpha}(p^2-m_0^2-m_5^2)^{-3}\nonumber\\
&\quad - \frac{1}{12}R(\pslash-\widetilde{m}_\psi)(p^2-m_0^2-m_5^2)^{-2}\nonumber\\
&\quad+\frac{1}{2}R_{\mu\lambda}p^\mu(\pslash-\widetilde{m}_\psi)\gamma^{\prime\lambda}(\pslash-\widetilde{m}_\psi)(p^2-m_0^2-m_5^2)^{-3}.\label{4.19}
\end{align}

It can be observed that for large $p$ the flat spacetime Green function behaves like $p^{-1}$ as usual, whereas the next order term $S_1$ behaves like $p^{-3}$. Higher order terms in the expansion \eqref{4.16} fall off even faster at large $p$ and cannot lead to pole terms in dimensional regularization for the terms that we are considering. 

If we return to \eqref{4.8a} we have
\begin{align}
\Gamma_{\textrm{D}\,2}^{(1)}&=-i\,{\textrm{Tr}}(X),\nonumber\\
&=\int dv_x \,\varphi(x)\,{\textrm{tr}}\lbrack S(x,x)(w+iw_5\gamma_5)\rbrack.\label{4.20}
\end{align}
We can use \eqref{4.16} to find the pole part of this result:
\begin{equation}
{\textrm{PP}}\Big\lbrace S(x,x)\Big\rbrace = {\textrm{PP}}\Big\lbrace \int\frac{d^np}{(2\pi)^n}\,\left\lbrack S_0(p) + S_1(p;x)\right\rbrack\Big\rbrace.\label{4.21}
\end{equation}
Terms which are odd in $p$ will integrate to zero, and terms which contain an odd number of $\gamma$-matrices will have a vanishing trace when this expression is substituted back into \eqref{4.20}. This reduces the number of terms that need to be retained considerably. (Note that although the next order term in the expansion \eqref{4.16} behaves like $p^{-4}$ by power counting, it vanishes due to odd powers of momentum or else odd numbers of $\gamma$-matrices when used in \eqref{4.21}.) After some calculation it can be shown that
\begin{equation}
{\textrm{PP}}\Big\lbrace{\textrm{tr}}\lbrack S(x,x)(w+iw_5\gamma_5)\rbrack\Big\rbrace = \frac{i}{2\pi^2\epsilon}(wm_0+w_5m_5)(m_0^2+m_5^2)+\frac{i}{24\pi^2\epsilon}(wm_0+w_5m_5)R.\label{4.22}
\end{equation}
The first term on the right hand side arises from $S_0$ and is just the result that would be found in flat spacetime; the second term comes from $S_1$. In obtaining this we have used the standard integrals of dimensional regularization~\cite{tHooftandVeltman}. Because we are only interested in the pole terms it is just as simple to expand the integrands in powers of $p$ and keep only the terms that behave like $p^{-4}$ and use
\begin{subequations}\label{4.23}
\begin{align}
{\textrm{PP}}\Big\lbrace\int\frac{d^np}{(2\pi)^n}\,\frac{1}{p^4}\Big\rbrace&=-\,\frac{i}{8\pi^2\epsilon},\label{4.23a}\\
{\textrm{PP}}\Big\lbrace\int\frac{d^np}{(2\pi)^n}\,\frac{p_\mu p_\nu}{p^6}\Big\rbrace&=-\,\frac{i}{32\pi^2\epsilon}\,\eta_{\mu\nu}.\label{4.23b}
\end{align}
\end{subequations}

We then find that 
\begin{equation}
{\textrm{PP}}\Big\lbrace\Gamma_{\textrm{D}\,2}^{(1)}\Big\rbrace =\frac{(wm_0+w_5m_5)}{2\pi^2\epsilon}\int dv_x\,\varphi(x)\Big(  m_0^2+m_5^2+\frac{1}{12}R\Big).\label{4.24}
\end{equation}
This result agrees with the corresponding terms in \eqref{3.14} that are linear in $\varphi$.

Turning next to \eqref{4.8b} we have
\begin{equation}
\Gamma_{\textrm{D}\,2}^{(2)}=\frac{i}{2}\int dv_x\,dv_{x^\prime}\,\varphi(x)\varphi(x^\prime)\Sigma_2(x,x^\prime),\label{4.25}
\end{equation}
where
\begin{equation}
\Sigma_2(x,x^\prime)={\textrm{tr}}\lbrack S(x,x^\prime)(w+iw_5\gamma_5)S(x^\prime,x)(w+iw_5\gamma_5)\rbrack.\label{4.26}
\end{equation}
If we denote the Fourier transform of $S(x,x^\prime)$ by $S(p;x^\prime)$ where $S(p;x^\prime)$ admits the expansion given in \eqref{4.16} we can write
\begin{equation}
\Sigma_2(x,x^\prime)=\int\frac{d^np}{(2\pi)^n}e^{ip_\mu y^\mu}\Sigma_2(p;x^\prime),\label{4.27}
\end{equation}
where
\begin{equation}
\Sigma_2(p;x^\prime)=\int\frac{d^nq}{(2\pi)^n}{\textrm{tr}}\lbrack S(p+q;x^\prime)(w+iw_5\gamma_5)S(q;x^\prime)(w+iw_5\gamma_5)\rbrack.\label{4.28}
\end{equation}
Using the asymptotic expansion in \eqref{4.16} and keeping only those terms that can lead to poles in dimensional regularization leads to
\begin{align}
{\textrm{PP}}\Big\lbrace\Sigma_2(p;x^\prime)\Big\rbrace&={\textrm{PP}}\Big\lbrace\int\frac{d^nq}{(2\pi)^n}\big\lbrace{\textrm{tr}}\lbrack S_0(p+q)(w+iw_5\gamma_5)S_0(q)(w+iw_5\gamma_5)\rbrack\nonumber\\
&+2\, {\textrm{tr}}\lbrack S_0(p+q)(w+iw_5\gamma_5)S_1(q;x^\prime)(w+iw_5\gamma_5)\rbrack\big\rbrace\Big\rbrace.\label{4.29}
\end{align}
We now use \eqref{4.17} and \eqref{4.19} along with \eqref{4.23} to find after some calculation
\begin{align}
{\textrm{PP}}\Big\lbrace\Sigma_2(p;x^\prime)\Big\rbrace&=\frac{i}{8\pi^2\epsilon}(w^2+w_5^2)p^2 - \frac{i}{2\pi^2\epsilon}\Big\lbrack (3w^2+w_{5}^2)m_0^2 + (w^2+3w_{5}^2)m_5^2\nonumber\\
&\qquad +4ww_5m_0m_5\Big\rbrack + \frac{i}{24\pi^2\epsilon}(w^2+w_5^2)R.\label{4.30}
\end{align}
The first two terms are those that would be found in the flat spacetime calculation. The term in $R$ arises solely from the second term in \eqref{4.29}. Using \eqref{4.30} in \eqref{4.27} results in
\begin{align}
{\textrm{PP}}\Big\lbrace\Sigma_2(x,x^\prime)\Big\rbrace&=\Big\lbrace-\frac{i}{8\pi^2\epsilon}(w^2+w_5^2)\Box_y - \frac{i}{2\pi^2\epsilon}\Big\lbrack (3w^2+w_{5}^2)m_0^2 + (w^2+3w_{5}^2)m_5^2\nonumber\\
&\qquad +4ww_5m_0m_5\Big\rbrack + \frac{i}{24\pi^2\epsilon}(w^2+w_5^2)R\Big\rbrack\delta(y)\label{4.31}
\end{align}
as the normal coordinate expression. In returning to general coordinates we must use the relation
\begin{equation}
\left(\Box_x+\frac{1}{3}R\right)\delta(x,x^\prime)=\Box_y\,\delta(y).\label{4.32}
\end{equation}
This result can be established in a similar manner to a related result in \cite{ParkerTomsbook}. From \eqref{4.25} after an integration by parts we find
\begin{align}
\Gamma_{\textrm{D}\,2}^{(2)}&=\int dv_x\,\Big\lbrace -\frac{1}{8\pi^2\epsilon}(w^2+w_5^2)\partial_\mu\varphi\partial^\mu\varphi + \frac{1}{4\pi^2\epsilon}\Big\lbrack (3w^2+w_{5}^2)m_0^2 + (w^2+3w_{5}^2)m_5^2\nonumber\\
&\qquad +4ww_5m_0m_5\Big\rbrack\varphi^2+\frac{1}{48\pi^2\epsilon} (w^2+w_5^2)R\varphi^2\Big\rbrace.\label{4.33}
\end{align}
This agrees with the terms in \eqref{3.14} that are quadratic in the background scalar field.

From \eqref{4.8c} and \eqref{4.5} we have
\begin{equation}
\Gamma_{\textrm{D}\,3}^{(1)}=-\frac{i}{3}\int dv_xdv_{x^\prime}dv_{x^{\prime\prime}}\varphi(x)\varphi(x^\prime)\varphi(x^{\prime\prime})\Sigma_3(x,x^\prime,x^{\prime\prime}),\label{4.34}
\end{equation}
where
\begin{equation}
\Sigma_3(x,x^\prime,x^{\prime\prime})={\textrm{tr}}\lbrack S(x,x^\prime)(w+iw_5\gamma_5) S(x^{\prime},x^{\prime\prime})(w+iw_5\gamma_5) S(x^{\prime\prime},x)(w+iw_5\gamma_5)\rbrack.\label{4.35}
\end{equation}
Power counting shows that only the first term, the flat spacetime part, of \eqref{4.16} contributes to the pole part of $\Sigma_3(x,x^\prime,x^{\prime\prime})$. A short calculation leads to the result
\begin{equation}
{\textrm{PP}}\Big\lbrace\Sigma_3(x,x^\prime,x^{\prime\prime})\Big\rbrace=\frac{3i}{2\pi^2\epsilon}(wm_0+w_5m_5)(w^2+w_5^2)\,\delta(x,x^\prime)\delta(x,x^{\prime\prime}).\label{4.36}
\end{equation}
When used back in \eqref{4.34} it is found that
\begin{equation}
{\textrm{PP}}\Big\lbrace\Gamma_{\textrm{D}\,3}^{(1)}\Big\rbrace=\frac{1}{2\pi^2\epsilon}(wm_0+w_5m_5)(w^2+w_5^2) \int dv_x\varphi^3(x),\label{4.37}
\end{equation}
a result that agrees with the cubic terms in $\varphi$ found in \eqref{3.14}.

Finally, from \eqref{4.8d} and \eqref{4.5} we have
\begin{equation}
\Gamma_{\textrm{D}\,4}^{(1)}=\frac{i}{4}\int dv_xdv_{x^\prime}dv_{x^{\prime\prime}}dv_{x^{\prime\prime\prime}}\varphi(x)\varphi(x^\prime)\varphi(x^{\prime\prime})\varphi(x^{\prime\prime\prime})\Sigma_4(x,x^\prime,x^{\prime\prime},x^{\prime\prime\prime}),\label{4.38}
\end{equation}
where
\begin{align}
\Sigma_4(x,x^\prime,x^{\prime\prime},x^{\prime\prime\prime})&={\textrm{tr}}\lbrack S(x,x^\prime)(w+iw_5\gamma_5) S(x^{\prime},x^{\prime\prime})(w+iw_5\gamma_5)\nonumber\\
&\quad\times S(x^{\prime\prime},x^{\prime\prime\prime})(w+iw_5\gamma_5) S(x^{\prime\prime\prime},x)(w+iw_5\gamma_5)\rbrack.\label{4.39}
\end{align}
For the pole part of this expression we may take $S_0(p)=\pslash/p^2$ and it is simple to show that
\begin{equation}
{\textrm{PP}}\Big\lbrace\Sigma_4(x,x^\prime,x^{\prime\prime},x^{\prime\prime\prime})\Big\rbrace=-\frac{i}{2\pi^2\epsilon}(w^2+w_5^2)^2\,\delta(x,x^\prime)\delta(x,x^{\prime\prime})\delta(x,x^{\prime\prime\prime}).\label{4.40}
\end{equation}
When used back in \eqref{4.38} this gives a result that agrees with the quartic terms in $\varphi$ in \eqref{3.14}.

We have therefore derived the pole terms found using the heat kernel method given in \eqref{3.14} in a completely independent way by using the local momentum space approach. An advantage of the local momentum space approach is that it makes the connection with traditional Feynman diagrams more transparent; the diagrams are the same as those found in flat spacetime, only the propagators have some additional curvature dependent terms.

\subsection{Yukawa coupling pole term}\label{sec4.2}

Up to now we have set the background Dirac field to zero. We will now determine the pole part of the terms in the effective action that depend on the background spinor field. This will enable the determination of the field renormalization for the Dirac field, the renormalization of the mass term, and the renormalization of the Yukawa coupling. 

Set
\begin{subequations}\label{4.2.1}
\begin{align}
\Psi&=\psi_q+\psi,\label{4.2.1a}\\
\Phi&=\varphi_q+\varphi,\label{4.2.1b}
\end{align}
\end{subequations}
where $\psi_q$ and $\varphi_q$ are the quantum fields that are integrated over in the functional integral and $\psi$ and $\varphi$ are the background fields. The terms in the classical action \eqref{2.1} and \eqref{2.2} that are quadratic in the quantum fields and so contribute to the one-loop effective action are
\begin{equation}
S_{\textrm{quad}}=\int dv_x\Big\lbrace \bar{\psi}_q(i\nablaslash-M)\psi_q - J_s\varphi_q
+\varphi_q\Big(-\Box-m_s^2-\xi R-\frac{\lambda}{2}\varphi^2-\eta\varphi\Big)\varphi_q\Big\rbrace,\label{4.2.3}
\end{equation}
where
\begin{align}
M&=m_0+im_5\gamma_5+\varphi(w+iw_5\gamma_5),\label{4.2.4}\\
J_s&=\bar{\psi}_q(w+iw_5\gamma_5)\psi+ \bar{\psi}(w+iw_5\gamma_5)\psi_q.\label{4.2.5}
\end{align}

The one-loop effective action is given by the functional integral
\begin{equation}
e^{i\,\Gamma^{(1)}}=\int\lbrack d\varphi_q\rbrack\lbrack d\psi_q\rbrack\,e^{i\,S_{\textrm{quad}} }.\label{4.2.6}
\end{equation}
We will perform the functional integration over the scalar field $\varphi_q$ first to obtain
\begin{equation}
e^{i\,\Gamma^{(1)}}=\int\lbrack d\psi_q\rbrack\,e^{\frac{i}{2}J_sGJ_s+i\,\int dv_x \bar{\psi}_q(i\nablaslash-M)\psi_q },\label{4.2.7}
\end{equation}
where we have abbreviated
\begin{equation}
J_sGJ_s=\int dv_x \int dv_{x^\prime}\,J_s(x)G(x,x^\prime)J_s(x^\prime),\label{4.2.8}
\end{equation}
with $G(x,x^\prime)$ the scalar field Green function given by
\begin{equation}
\Big(\Box+m_s^2+\xi R+\frac{\lambda}{2}\varphi^2+\eta\varphi\Big)G(x,x^\prime)=\delta(x,x^\prime).\label{4.2.9}
\end{equation}
We are only after the pole part of the one-loop effective action so that we can expand the exponential term $e^{\frac{i}{2}J_sGJ_s} $ in a Taylor series,
\begin{equation}
e^{\frac{i}{2}J_sGJ_s}=1+\frac{i}{2}J_sGJ_s+\cdots.\label{4.2.10}
\end{equation}
The first term in the expansion just gives rise to the second term of \eqref{2.3} that we have already calculated. The second term gives rise to pole terms in the effective action that are quadratic in the background Dirac field. Higher order terms in the expansion are finite in dimensional regularization so cannot contain pole terms. The one-loop effective action in \eqref{4.2.7} can be expressed as 
\begin{equation}
\Gamma^{(1)}=-i\left\langle e^{\frac{i}{2}J_sGJ_s}\right\rangle,\label{4.2.11}
\end{equation}
where $\langle \cdots\rangle$ means to evaluate using Wick's theorem keeping only terms that correspond to one-particle irreducible Feynman diagrams. With our conventions we have the fundamental correlation function
\begin{equation}
\langle \psi_{q\,\alpha}(x)\bar{\psi}_{q\,\beta}(x^\prime)\rangle=-i\,\Delta_{\alpha\beta}(x,x^\prime),\label{4.2.12}
\end{equation}
with $\alpha$ and $\beta$ spinor indices, and $\Delta(x,x^\prime)$ the Dirac Green's function defined by
\begin{equation}
(i\nablaslash-M)\Delta(x,x^\prime)=-I\delta(x,x^\prime).\label{4.2.13}
\end{equation}
Comparison with \eqref{4.3} shows that when we set $\varphi=0$ we have $\Delta(x,x^\prime)=S(x,x^\prime)$.

It is now straightforward to show that if we ignore the terms in the effective action that are independent of the background Dirac field that we have calculated in the earlier sections we have
\begin{align}
{\textrm{PP}}\Big\lbrace \Gamma^{(1)}\Big\rbrace &={\textrm{PP}}\Big\lbrace \frac{1}{2}\langle J_sGJ_s\rangle\Big\rbrace\nonumber\\
&=-i\,{\textrm{PP}}\Big\lbrace\int dv_xdv_{x^\prime}G(x,x^\prime)\bar{\psi}(x)(w+iw_5\gamma_5)\Delta(x,x^\prime) (w+iw_5\gamma_5) \psi(x^\prime)\Big\rbrace.\label{4.2.14}
\end{align}
If we use the local momentum space expansion for both $G(x,x^\prime)$ and $\Delta(x,x^\prime)$ it follows that
\begin{equation}
G(x,x^\prime)\,\Delta(x,x^\prime)=\int\frac{d^np}{(2\pi)^n}\,e^{ip_\mu y^\mu}\,Y(p;x^\prime),\label{4.2.15}
\end{equation}
where
\begin{equation}
Y(p;x^\prime)=\int\frac{d^nq}{(2\pi)^n}\,G(p-q;x^\prime)\Delta(q;x^\prime).\label{4.2.16}
\end{equation}
The pole terms in this expression will only involve terms that behave like $q^{-4}$ when expanded for large $q$.  Even though $G(p-q;x^\prime)$ and $\Delta(q;x^\prime)$ have complicated expansions in powers of curvature the leading order pole term can be seen to involve only the flat spacetime parts of the Green functions. The leading order terms that can give rise to a pole are found to be
\begin{equation}
G(p-q;x^\prime)\Delta(q;x^\prime)=-\frac{\qslash}{q^4}+\frac{\qslash M\qslash}{q^6}-\frac{2p_\mu q^\mu\,\qslash}{q^6}+\cdots.\label{4.2.17}
\end{equation}
The first term here is odd in $q$ and so will integrate to zero. The leading behaviour is seen to be $q^{-4}$. (This is how it is easy to see that higher order terms in the expansion \eqref{4.2.10} will not lead to any pole; the analogous integrands fall off faster than $q^{-4}$ for large $q$.) Using the basic integrals in \eqref{4.23} it can be seen that
\begin{align}
{\textrm{PP}}\Big\lbrace Y(p;x^\prime)\Big\rbrace&=-\frac{i}{32\pi^2\epsilon}\left(\gamma_\mu\,M\gamma^\mu-2p_\mu\gamma^\mu\right)\nonumber\\
&=-\frac{i}{16\pi^2\epsilon}\left\lbrack 2(m_0+w\varphi)-2i(m_5+w_5\varphi)\gamma_5-p_\mu\gamma^\mu\right\rbrack.\label{4.2.18}
\end{align}
When used in \eqref{4.2.15}, after returning from Riemann normal coordinates to general coordinates, we have
\begin{equation}
{\textrm{PP}}\Big\lbrace G(x,x^\prime)\,\Delta(x,x^\prime)\Big\rbrace=-\frac{i}{16\pi^2\epsilon}\left\lbrack 2(m_0+w\varphi)-2i(m_5+w_5\varphi)\gamma_5+i\nablaslash\right\rbrack\delta(x,x^\prime).\label{4.2.19}
\end{equation}
After a bit of algebra we find from \eqref{4.2.14} that
\begin{align}
{\textrm{PP}}\Big\lbrace \Gamma^{(1)}\Big\rbrace &=-\frac{1}{16\pi^2\epsilon}\int dv_x\Big\lbrace i(w^2+w_5^2)\bar{\psi}\nablaslash\psi + 2(w^2+w_5^2)\,\varphi\,\bar{\psi}(w+iw_5\gamma_5)\psi\nonumber\\
&+2\lbrack(w^2-w_5^2)m_0+2ww_5m_5\rbrack\bar{\psi}\psi -2i\lbrack(w^2-w_5^2)m_5-2ww_5m_0\rbrack\bar{\psi}\gamma_5\psi\Big\rbrace.\label{4.2.20}
\end{align}
When this result is combined with the results of the previous sections it will give us the complete set of pole terms for the one-loop effective action. The evaluation of the necessary counterterms will be given in the next section.

\section{Renormalization counterterterms and renormalization group functions}\label{sec5}

\subsection{Counterterms}\label{sec5.1}

In addition to the one-loop contributions to the effective action that we have found there is also the contribution from the classical action \eqref{2.1} with \eqref{2.2}. We set the fields equal to their bare background values $\varphi_{\bare}$ and $\psi_{\bare}$. These fields can be expressed in terms of their renormalized values $\varphi$ and $\psi$ by introducing field renormalization factors and the unit of mass $\mu$. Following `t Hooft~\cite{tHooft1973} the unit of mass is used to ensure that all renormalized quantities have their dimensions fixed at the values in four spacetime dimensions for any value of the spacetime dimension $n=4+\epsilon$. Similar considerations apply to all of the bare coupling constants in \eqref{2.1} and \eqref{2.2}. We will define the necessary counterterms as follows:
\begin{subequations}\label{5.1}
\begin{align}
\varphi_\bare&=\mu^{\epsilon/2}(1+\delta Z_\varphi)\varphi,\label{5.1a}\\
m_{s\,\bare}^2&=m_s^2+\delta m_s^2,\label{5.1b}\\
\xi_\bare&=\xi+\delta\xi,\label{5.1c}\\
\lambda_\bare&=\mu^{-\epsilon}(\lambda+\delta\lambda),\label{5.1d}\\
\eta_\bare&=\mu^{-\epsilon/2}(\eta+\delta\eta),\label{5.1e}\\
\tau_\bare&=\mu^{\epsilon/2}(\tau+\delta\tau),\label{5.1f}\\
\gamma_\bare&=\mu^{\epsilon/2}(\gamma+\delta\gamma),\label{5.1g}\\
\psi_\bare&=\mu^{\epsilon/2}(1+\delta Z_\psi)\psi,\label{5.1h}\\
m_{0\,\bare}&=m_0+\delta m_0,\label{5.1i}\\
m_{5\,\bare}&=m_5+\delta m_5,\label{5.1j}\\
w_{0\,\bare}&=\mu^{-\epsilon/2}(w_0+\delta w_0),\label{5.1k}\\
w_{5\,\bare}&=\mu^{-\epsilon/2}(w_5+\delta w_5),\label{5.1l}\\
\Lambda_{\bare}&=\mu^{\epsilon}(\Lambda+\delta\Lambda),\label{5.1m}\\
\kappa_{\bare}&=\mu^{\epsilon}(\kappa+\delta\kappa),\label{5.1n}\\
\alpha_{i\,\bare}&=\mu^{\epsilon}(\alpha_i+\delta\alpha_i),\ i=1,2,3.\label{5.1o}
\end{align}
\end{subequations}
All of the counterterms can be expanded in a series of poles and number of loops. It is easy to see (using $d^nx=\mu^{-\epsilon}d^4x$) that the counterterm part of the scalar action \eqref{2.2a} is (where $dv_x=|\det(g_{\mu\nu})|^{1/2}\,d^4x$)
\begin{align}
S_{\textrm{scalar}}^{\textrm{ct}}&=\int dv_x\Big\lbrack \delta Z_{\varphi}\partial^\mu\varphi\partial_\mu\varphi-\Big(\frac{1}{2}\delta m_s^2+m_s^2\delta Z_{\varphi}\Big)\varphi^2  -\Big(\frac{1}{2}\delta \xi+\xi\delta Z_{\varphi}\Big)R\,\varphi^2\nonumber\\
&-\Big(\frac{\delta\lambda}{4!}+\frac{\lambda}{3!}\delta Z_{\varphi}\Big)\varphi^4 -\Big(\frac{\delta\eta}{3!}+\frac{\eta}{2}\delta Z_{\varphi}\Big)\varphi^3-\Big(\delta\tau+\tau\delta Z_{\varphi}\Big)\varphi -\Big(\delta\gamma+\gamma\delta Z_{\varphi}\Big)R\,\varphi\Big\rbrack.\label{5.2}
\end{align}
The counterterm part of the Dirac action in \eqref{2.2b} is
\begin{align}
S_{\textrm{spinor}}^{\textrm{ct}}&=\int dv_x\Big\lbrack 2i\,\delta Z_\psi\,\bar{\psi}\nablaslash\psi-(\delta m_0+2m_0\,\delta Z_\psi)\,\bar{\psi}\psi-i\,(\delta m_5+2m_5\,\delta Z_\psi)\,\bar{\psi}\gamma_5\psi\nonumber\\
&-(\delta w+w\,\delta Z_\varphi+2w\,\delta Z_\psi)\,\varphi\bar{\psi}\psi -i\,(\delta w_5+w_5\,\delta Z_\varphi+2w_5\,\delta Z_\psi)\,\varphi\bar{\psi}\gamma_5\psi\Big\rbrack.\label{5.3}
\end{align}
The counterterm part of the gravitational action \eqref{2.2c} is
\begin{align}
S_{\textrm{grav}}^{\textrm{ct}}&=\int dv_x\Big(\delta\Lambda+\delta\kappa\,R+\delta\alpha_{1}\,R^{\mu\nu\lambda\sigma}R_{\mu\nu\lambda\sigma} + \delta\alpha_{2}\,R^{\mu\nu}R_{\mu\nu} + \delta\alpha_{3}\,R^2 \Big).\label{5.4}
\end{align}

We require that the counterterms in \eqref{5.2}--\eqref{5.4} cancel the poles that have been found in the one-loop effective action given by combining \eqref{3.14} and \eqref{4.2.20}. This fixes
\begin{subequations}\label{5.5}
\begin{align}
\delta Z_\varphi&=\frac{1}{8\pi^2\epsilon}\,(w^2+w_5^2),\label{5.5a}\\
\delta m_s^2&=-\frac{1}{16\pi^2\epsilon}\,\Big\lbrack\eta^2+(\lambda+4w^2+4w_5^2)m_s^2 -8(3w^2+w_5^2)m_0^2\nonumber\\
&\qquad-32ww_5m_0m_5-8(3w_5^2+w^2)m_5^2\Big\rbrack,\label{5.5b}\\
\delta\xi&=-\frac{1}{16\pi^2\epsilon}\,\Big\lbrack\lambda+4\,(w^2+w_5^2)\Big\rbrack\,\Big(\xi-\frac{1}{6}\Big),\label{5.5c}\\
\delta\lambda&=-\frac{1}{16\pi^2\epsilon}\,\Big\lbrack3\lambda^2-48(w^2+w_5^2)^2+8(w^2+w_5^2)\,\lambda\Big\rbrack,\label{5.5d}\\
\delta\eta&=-\frac{1}{16\pi^2\epsilon}\,\Big\lbrack3\lambda\eta-48(w^2+w_5^2)(wm_0+w_5m_5)+6(w^2+w_5^2)\,\eta \Big\rbrack,\label{5.5e}\\
\delta\tau&=-\frac{1}{16\pi^2\epsilon}\,\Big\lbrack\eta m_s^2-8(m_0^2+m_5^2)(wm_0+w_5m_5)+2(w^2+w_5^2)\,\tau\Big\rbrack,\label{5.5f}\\
\delta\gamma&=-\frac{1}{16\pi^2\epsilon}\,\Big\lbrack\Big(\xi-\frac{1}{6}\Big)\eta - \frac{2}{3}\,(wm_0+w_5m_5)+ 2\,(w^2+w_5^2)\,\gamma\Big\rbrack,\label{5.5g}\\
\delta Z_\psi&=\frac{1}{32\pi^2\epsilon}\,(w^2+w_5^2),\label{5.5h}\\
\delta m_0&=-\frac{1}{16\pi^2\epsilon}\,\Big\lbrack (3w^2-w_5^2)\,m_0+4ww_5m_5\Big\rbrack,\label{5.5i}\\
\delta m_5&=-\frac{1}{16\pi^2\epsilon}\,\Big\lbrack (3w_5^2-w^2)\,m_5+4ww_5m_0\Big\rbrack,\label{5.5j}\\
\delta w&=-\frac{5}{16\pi^2\epsilon}\,w\,(w^2+w_5^2),\label{5.5k}\\
\delta w_5&=-\frac{5}{16\pi^2\epsilon}\,w_5\,(w^2+w_5^2),\label{5.5l}\\
\delta \Lambda&=\frac{1}{32\pi^2\epsilon}\,\Big\lbrack m_s^4-4(m_0^2+m_5^2)^2\Big\rbrack,\label{5.5m}\\
\delta \kappa&=\frac{1}{48\pi^2\epsilon}\,\Big\lbrack 3\Big(\xi-\frac{1}{6}\Big)m_s^2-m_0^2-m_5^2\Big\rbrack,\label{5.5n}\\
\delta \alpha_1&=\frac{1}{640\pi^2\epsilon},\label{5.5o}\\
\delta \alpha_2&=\frac{1}{960\pi^2\epsilon},\label{5.5p}\\
\delta \alpha_3&=\frac{1}{32\pi^2\epsilon}\,\xi\Big(\xi-\frac{1}{3}\Big).\label{5.5q}
\end{align}
\end{subequations}

The results in \eqref{5.5} give the complete set of one-loop counterterms for the theory. Unlike the case of simple $\lambda\varphi^4$ theory a scalar field renormalization is necessary. It would not be possible to calculate this by assuming a constant background scalar field and its neglect would lead to incorrect counterterms and renormalization group functions. Even in cases where the cubic and linear terms are not present in the original scalar action the presence of the Yukawa interaction necessitates adding in cubic and linear counterterms to remove poles from the effective action as is evident from \eqref{5.5e}--\eqref{5.5g}. (This is also apparent from the traditional Feynman diagram approach as the Yukawa interaction will lead to divergent one-point and three-point functions.) Another feature worth noting is that nether the scalar field mass nor the fermion masses are multiplicatively renormalized. Finally we note that the quadratic non-minimal term $\xi$ appears in the combination $(\xi-1/6)$ everywhere except in the last counterterm \eqref{5.5q}. The reason for this exception is that the fermion generates a pole in $R^2$ that cannot involve $\xi$.

\subsection{Renormalization group functions}\label{sec5.2}

The renormalization group functions can be calculated from the counterterms in \eqref{5.5} as described originally by `t~Hooft~\cite{tHooft1973}. We will follow the curved spacetime treatment in \cite{ParkerTomsbook} except that we will use the more conventional unit of mass $\mu$ rather than the unit of length used in \cite{ParkerTomsbook}. 

If $q_{i\,\bare}$ represents any of the bare background fields or bare coupling constants then we have
\begin{equation}
q_{i\,\bare}=\mu^{-\epsilon\alpha_{q_i}}\,(q_i+\delta q_i).\label{5.2.1}
\end{equation}
Here $\alpha_{q_i}$ is just some number. See \eqref{5.1} for the explicit expressions for the theory in the present paper. Because bare quantities cannot depend on the renormalization mass $\mu$ it follows that if we change the renormalization mass then the renormalized quantities must change and it can be shown that
\begin{equation}
\mu\frac{d}{d\mu}\,q_i=\beta_{q_i}.\label{5.2.2}
\end{equation}
Here $\beta_{q_i}$ is the renormalization group function defined in terms of the coefficient of the simple pole in the counterterm. If we expand the counterterms as a series of poles,
\begin{equation}
\delta q_i=\frac{1}{\epsilon}\,\delta q_{i\,1}+\frac{1}{\epsilon^2}\,\delta q_{i\,2}+\cdots,\label{5.2.3}
\end{equation}
then
\begin{equation}
\beta_{q_i}=\alpha_{q_i}\,\delta q_{i\,1} - \sum_{j}\alpha_{q_j}\,q_j\,\frac{\partial}{\partial q_j}\,\delta q_{i\,1}.\label{5.2.4}
\end{equation}
Application of \eqref{5.2.4} to each of the quantities in \eqref{5.5} leads to the following renormalization group functions
\begin{subequations}\label{5.2.5}
\begin{align}
\beta_{m_s^2}&=\frac{1}{16\pi^2}\Big\lbrack \eta^2+(\lambda+4w^2+4w_5^2)m_s^2-8(3w^2+w_5^2)m_0^2 -8(w^2+3w_5^2)m_5^2 \nonumber\\
&\qquad -32ww_5m_0m_5\Big\rbrack,\label{5.2.5a}\\
\beta_{\xi}&=\frac{1}{16\pi^2}(\lambda+4w^2+4w_5^2)\left(\xi-\frac{1}{6}\right),\label{5.2.5b}\\
\beta_{\lambda}&=\frac{1}{16\pi^2}\Big\lbrack 3\lambda^2+8(w^2+w_5^2)\lambda -48(w^2+w_5^2)^2\Big\rbrack,\label{5.2.5c}\\
\beta_{\eta}&=\frac{1}{16\pi^2}\Big\lbrack 3\lambda\eta+6(w^2+w_5^2)\eta-48(w^2+w_5^2)(wm_0+w_5m_5)\Big\rbrack,\label{5.2.5d}\\
\beta_{\tau}&=\frac{1}{16\pi^2}\Big\lbrack \eta m_s^2-8(m_0^2+m_5^2)(wm_0+w_5m_5)+2(w^2+w_5^2)\tau\Big\rbrack,\label{5.2.5e}\\
\beta_{\gamma}&=\frac{1}{16\pi^2}\Big\lbrack \left(\xi-\frac{1}{6}\right)\eta - \frac{2}{3}(wm_0+w_5 m_5)+2(w^2+w_5^2)\gamma\Big\rbrack,\label{5.2.5f}\\
\beta_{m_0}&=\frac{1}{16\pi^2}\Big\lbrack (3w^2-w_5^2)m_0+4ww_5m_5\Big\rbrack,\label{5.2.5g}\\
\beta_{m_5}&=\frac{1}{16\pi^2}\Big\lbrack (3w_5^2-w^2)m_5+4ww_5m_5\Big\rbrack,\label{5.2.5h}\\
\beta_{w}&=\frac{5}{16\pi^2} w (w^2+w_5^2),\label{5.2.5i}\\
\beta_{w_5}&=\frac{5}{16\pi^2} w_5 (w^2+w_5^2),\label{5.2.5j}\\
\beta_{\Lambda}&=\frac{1}{32\pi^2}\Big\lbrack 4(m_0^2+m_5^2)^2-m_s^4\Big\rbrack,\label{5.2.5k}\\
\beta_{\kappa}&=\frac{1}{48\pi^2}\Big\lbrack m_0^2+m_5^2-3 \left(\xi-\frac{1}{6}\right)m_s^2\Big\rbrack,\label{5.2.5l}\\
\beta_{\alpha_1}&=- \frac{1}{640\pi^2},\label{5.2.5m}\\
\beta_{\alpha_2}&=- \frac{1}{960\pi^2},\label{5.2.5n}\\
\beta_{\alpha_3}&=- \frac{1}{32\pi^2}\,\xi\left(\xi-\frac{1}{3}\right) ,\label{5.2.5o}\\
\beta_{\varphi}&=- \frac{1}{8\pi^2}(w^2+w_5^2)\varphi,\label{5.2.5p}\\
\beta_{\psi}&=- \frac{1}{32\pi^2}(w^2+w_5^2)\psi.\label{5.2.5q}
\end{align}
\end{subequations}

The results of \eqref{5.2.5d}--\eqref{5.2.5f} show that we can only consistently set $\eta=\tau=\gamma=0$ in the case of massless fermions ($m_0=m_5=0$). \eqref{5.2.5b} shows that the conformal value $\xi=1/6$ is a fixed point to one-loop order, although this would not be expected to persist beyond one-loop.

It is possible to solve the renormalization group equations for $w,w_5,\lambda$ exactly. The fact that the third term in \eqref{5.2.5c} is negative means that it is possible for $\lambda$ to approach zero from its initial value. For this to happen the initial value of $\lambda$ must not be too large. If this is not the case then $\lambda$ increases as it does in normal quartic scalar field theory which invalidates perturbation theory at some point. Even if it is possible for $\lambda$ to decrease to zero this value is not a renormalization group fixed point. We present a brief analysis of the solution for the running value of $\lambda$ in the Appendix~\ref{append}.

\subsection{Effective potential}\label{sec5.3}

The seminal paper of Coleman and Weinberg~\cite{ColemanandWeinberg} showed how the renormalization group could be used to evaluate the effective potential in the case where there was no mass scale present in the classical theory. It was demonstrated how an exact result could be found for the effective potential in terms of the renormalization group functions. To one-loop order the result agrees with a direct evaluation of the effective potential using other means. The renormalization group was applied to the effective action in curved spacetime in \cite{TomsRG} and to the effective potential in \cite{BuchbinderandOdintsovRG} to discuss symmetry breaking in curved spacetime. The situation is reviewed in \cite{ParkerTomsbook} and we follow the general framework set out there except that we will use the more conventional unit of mass rather than length. We will also content ourselves with just the one-loop expressions rather than obtaining exact results in terms of the renormalization group functions and then expanding them to one-loop order. This makes the calculation very simple and straightforward.

We will set all dimensional coupling constants in the classical theory to zero. Specifically we take $m_s^2=\eta=\tau=\gamma=m_0=m_5=0$ in \eqref{2.2}. The renormalization group functions in \eqref{5.2.5} ensure that these dimensional parameters remain zero to one-loop order at least. Based on the form of the classical action in \eqref{2.2} we will consider terms in the renormalized effective action of the form
\begin{align}
\Gamma&=\int dv_x\Big\lbrack \frac{1}{2}Z(\varphi)\partial^\mu\varphi\partial_\mu\varphi-V_0(\varphi)-RV_1(\varphi)\nonumber\\
&\qquad+\alpha_{1}(\varphi)R^{\mu\nu\lambda\sigma}R_{\mu\nu\lambda\sigma} + \alpha_{2}(\varphi)R^{\mu\nu}R_{\mu\nu} + \alpha_{3}(\varphi)R^2 \Big\rbrack,\label{5.3.1}
\end{align}
where $Z(\varphi),V_0(\varphi),V_1(\varphi),\alpha_i(\varphi)$ are to be determined. There will be other terms in the full one-loop effective action that cannot be determined by renormalization group methods; the expression in \eqref{5.3.1} can be viewed as the first few terms in the expansion of the effective action in powers of the curvature. The potentials $Z(\varphi),V_0(\varphi),V_1(\varphi),\alpha_i(\varphi)$ also depend on the renormalization scale $\mu$ as well as the coupling constants. Because there is no mass scale in the theory apart from $\mu$ and $\varphi$, as Coleman and Weinberg~\cite{ColemanandWeinberg} argued, these two quantities must occur in the dimensionless combination $\varphi/\mu$. We adopt the Coleman and Weinberg normalization conditions for $Z(\varphi)$ and $V_0(\varphi)$
\begin{align}
Z(\varphi=\mu)&=1,\label{5.3.2}\\
\left.\frac{\partial^4 V_0(\varphi)}{\partial\varphi^4}\right|_{\varphi=\mu}&=\lambda,\label{5.3.3}
\end{align}
along with analogous expressions for the curved spacetime terms
\begin{align}
\left.\frac{\partial^2 V_1(\varphi)}{\partial\varphi^2}\right|_{\varphi=\mu}&=\xi,\label{5.3.4}\\
\alpha_i(\varphi=\mu)&=\alpha_i,\ {\textrm{for}}\ i=1,2,3.\label{5.3.5}
\end{align}
The form of the renormalization group equations requires the $\varphi/\mu$ dependence to be in the form of $\ln(\varphi^2/\mu^2)$. We will therefore take
\begin{subequations}\label{5.2.6}
\begin{align}
Z(\varphi)&=1+A\,\ln(\varphi^2/\mu^2),\label{5.3.6a}\\
V_0(\varphi)&=\frac{\lambda}{4!}\,\varphi^4+B\,\varphi^4\Big\lbrack \ln(\varphi^2/\mu^2)-\frac{25}{6}\Big\rbrack,\label{5.3.6b}\\
V_1(\varphi)&=\frac{1}{2}\,\xi\,\varphi^2+C\,\varphi^2\lbrack \ln(\varphi^2/\mu^2)-3\rbrack,\label{5.3.6c}\\
\alpha_i(\varphi)&=\alpha_i+D_i\, \ln(\varphi^2/\mu^2).\label{5.3.6d}
\end{align}
\end{subequations}
Here $A,B,C$ and $D_i$ depend only on the coupling constants $\lambda,\xi,w,w_5$ and are to be determined.

When we change the renormalization scale from $\mu$ to $\mu^\prime$ the terms in the effective action \eqref{5.3.1} cannot change. The field $\varphi$ as well as the coupling constants will all change in accordance with their renormalization group equations the general form of which is \eqref{5.2.2}. To one-loop order we have (with $q_i^\prime=q_i(\mu^\prime)$)
\begin{subequations}
\begin{align}
\varphi^\prime&=\big\lbrack 1+\tilde{\beta}_{\varphi}\,\ln(\mu^\prime/\mu)\big\rbrack\varphi,\label{5.3.7a}\\
\lambda^\prime&=\lambda+\beta_\lambda\ln(\mu^\prime/\mu),\label{5.3.7b}\\
\xi^\prime&=\lambda+\beta_\xi\ln(\mu^\prime/\mu),\label{5.3.7c}\\
\alpha_i^\prime&=\alpha_i+\beta_{\alpha_i}\ln(\mu^\prime/\mu),\label{5.3.7d}
\end{align}
\end{subequations}
where the one-loop expressions for the renormalization group functions was given by \eqref{5.2.5} and we have defined $\beta_\varphi=\tilde{\beta}_{\varphi}\varphi$ to explicitly remove the $\varphi$ dependence in the following.

Requiring invariance of the derivative term in \eqref{5.3.1} under the change $\mu\rightarrow\mu^\prime$ results in 
\begin{align}
A&=\tilde{\beta}_{\varphi}\nonumber\\
&=- \frac{1}{8\pi^2}(w^2+w_5^2),\label{5.3.8}
\end{align}
when we use \eqref{5.3.6a} along with \eqref{5.3.7a} and \eqref{5.2.5p} and work consistently to one-loop order. 

Invariance of the term in $V_0$ in \eqref{5.3.1} under the change $\mu\rightarrow\mu^\prime$ results in 
\begin{align}
B&=\frac{1}{48}\,\beta_\lambda+\frac{\lambda}{12}\,\tilde{\beta}_{\varphi}\nonumber\\
&= \frac{1}{256\pi^2}\lbrack\,3\lambda^2-16\,(w^2+w_5^2)^2\rbrack,\label{5.3.9}
\end{align}
when we use \eqref{5.3.6b} along with \eqref{5.3.7a},\eqref{5.3.7b},\eqref{5.2.5p} and \eqref{5.2.5c}.

Invariance of the term in $V_1$ in \eqref{5.3.1} under the change $\mu\rightarrow\mu^\prime$ results in 
\begin{align}
C&=\frac{1}{4}\,\beta_\xi+\frac{1}{2}\,\xi\,\tilde{\beta}_{\varphi}\nonumber\\
&= \frac{1}{192\pi^2}\lbrack\,3\lambda\big(\xi-1/6\big)-2\,(w^2+w_5^2)\rbrack,\label{5.3.10}
\end{align}
when we use \eqref{5.3.6c} along with \eqref{5.3.7a},\eqref{5.3.7c},\eqref{5.2.5p} and \eqref{5.2.5b}. It can be noted that this coefficient does not vanish even if $\xi$ takes the one-loop fixed point conformal value of $1/6$.

Invariance of the quadratic curvature terms in \eqref{5.3.1} under the change $\mu\rightarrow\mu^\prime$ results in $D_i=\frac{1}{2}\beta_{\alpha_i}$ using \eqref{5.3.6d} and \eqref{5.3.7d}. The explicit results for $\alpha_i(\varphi)$ defined by \eqref{5.3.7d} are
\begin{align}
\alpha_1(\varphi)&=\alpha_1-\frac{1}{1280\pi^2}\,\ln(\varphi^2/\mu^2),\label{5.3.11a}\\
\alpha_2(\varphi)&=\alpha_2-\frac{1}{1920\pi^2}\,\ln(\varphi^2/\mu^2),\label{5.3.11b}\\
\alpha_3(\varphi)&=\alpha_1-\frac{1}{64\pi^2}\,\xi(\xi-1/3)\,\ln(\varphi^2/\mu^2),\label{5.3.11c}
\end{align}
using \eqref{5.2.5m}--\eqref{5.2.5o}.

This completes the evaluation of the terms in the one-loop effective action \eqref{5.3.1} that can be determined by renormalization group considerations. It is worth pointing out that the inclusion of the proper field renormalization is crucial for obtaining the correct effective potential and the field renormalization cannot be found by expanding about a constant background field.

\section{Field redefinitions and an anomaly}\label{secanomaly}

As we mentioned in Sec.~\ref{sec2} it is possible to perform a chiral transformation on the spinor field in \eqref{2.2b} to remove the pseudoscalar mass term in $m_5$ from the classical action and leave only the usual mass term in $m_0$. Let (leaving off the subscript `B')
\begin{equation}
\Psi(x)=e^{-i\theta\gamma_5}\,\Psi^\prime(x)\label{FD1}
\end{equation}
for constant $\theta$. Then
\begin{equation}
\bar{\Psi}(x)=\bar{\Psi}^\prime(x)\,e^{-i\theta\gamma_5},\label{FD2}
\end{equation}
and we have
\begin{equation}
\bar{\Psi}(x)(m_0+im_5\gamma_5)\Psi(x)=\bar{\Psi}^\prime(x)(m_{0}^{\prime} + im_{5}^{\prime}\gamma_5) \Psi^\prime(x),\label{FD3}
\end{equation}
where
\begin{subequations}\label{FD4}
\begin{align}
m_{0}^{\prime}&=m_0\,\cos(2\theta)+m_5\,\sin(2\theta),\label{FD4a}\\
m_{5}^{\prime}&=m_5\,\cos(2\theta)-m_0\,\sin(2\theta).\label{FD4b}
\end{align}
\end{subequations}
If we choose
\begin{subequations}\label{FD5}
\begin{align}
\sin(2\theta)&=\frac{m_5}{(m_0^2+m_5^2)^{1/2}},\label{FD5a}\\
\cos(2\theta)&=\frac{m_0}{(m_0^2+m_5^2)^{1/2}},\label{FD5b}
\end{align}
\end{subequations}
then from \eqref{FD4} we have
\begin{subequations}\label{FD6}
\begin{align}
m_{0}^{\prime}&=(m_0^2+m_5^2)^{1/2},\label{FD6a}\\
m_{5}^{\prime}&=0.\label{FD6b}
\end{align}
\end{subequations}
This leaves only the normal mass term present in the spinor part of the action. (We could equally well choose $\theta$ so that $m_{0}^\prime=0$ and retain just the pseudoscalar mass term if we choose.)

The Yukawa coupling term will also change under the transformation in \eqref{FD1} and \eqref{FD2}. We will end up with the fermion part of the action in \eqref{2.2b} becoming (again dropping the subscript `B')
\begin{equation}
S_{\textrm{spinor}}^\prime=\int dv_x\Big\lbrack \bar{\Psi}^\prime(i\gamma^\mu\nabla_\mu-m_{0}^\prime)\Psi-\Phi\bar{\Psi}(w^\prime+iw_{5}^\prime\gamma_5)\Psi^\prime\big\rbrack,\label{FD7}
\end{equation}
where
\begin{subequations}\label{FD8}
\begin{align}
w^{\prime}&=\frac{wm_0+w_5m_5}{(m_0^2+m_5^2)^{1/2}},\label{FD8a}\\
w_{5}^{\prime}&=\frac{w_5m_0-wm_5}{(m_0^2+m_5^2)^{1/2}}.\label{FD8b}
\end{align}
\end{subequations}
Although the classical theories based on \eqref{2.2b} and \eqref{FD7} should be identical, it is not obvious that this will also be true for the quantum theories. The equivalence under the chiral transformation \eqref{FD1} may not hold at the quantum level. An almost identical situation holds for the axial vector current with the difference being that $\theta$ is not assumed to be constant. 

In order to study the quantum equivalence of the theories based on \eqref{2.2b} and \eqref{FD7} we will study the behaviour of the functional measure based in the pioneering work of Fujikawa~\cite{Fujikawa1,Fujikawa2,Fujikawa3}. Our notation and approach follows that in \cite[Sec.~5.9]{ParkerTomsbook}. We will only consider the spinor part of the effective action here as the Bose part is not affected by the chiral transformation \eqref{FD1}. If we call the effective action based on \eqref{2.2b} $\Gamma$ and that based on \eqref{FD7} $\Gamma^\prime$ then the change of variables in \eqref{FD1} and \eqref{FD2} results in the formal change in the functional measure being
\begin{equation}
\lbrack d\Psi d\bar{\Psi}\rbrack=(\det\,C_{NN^\prime})^{-2}\,\lbrack d\Psi^\prime d\bar{\Psi}^\prime\rbrack,\label{FD9}
\end{equation}
where the inverse Jacobian arises from the nature of the Grassmann variables. The result for $C_{NN^\prime}$ is
\begin{equation}
C_{NN^\prime}=\mu\int dv_x\bar{\psi}_N(x)\,e^{i\theta\gamma_5}\,\psi_{N^\prime}(x),\label{FD10}
\end{equation}
where $\lbrace \psi_N(x)\rbrace$ is a complete orthonormal set of solutions to
\begin{equation}
D\psi_N(x)=\lambda_N\psi_N(x),\label{FD11}
\end{equation}
and $\mu$ is an arbitrary unit of mass which keeps $C_{NN^\prime}$ dimensionless. In \eqref{FD11} $D$ is the Dirac operator that is found from \eqref{2.2b} and was given in \eqref{2.3}, or else that which is found from \eqref{FD7}. We will show that it does not matter which one is chosen. The relationship between $\Gamma$ and $\Gamma^\prime$ is then
\begin{equation}
\Gamma^\prime=\Gamma+2i\ln\det\,C_{NN^\prime}.\label{FD12}
\end{equation}

The result in \eqref{FD10} is the same as that found in the analysis of the chiral anomaly except that $\theta$ here is constant and is fixed by \eqref{FD5}. In the evaluation of the chiral anomaly $\theta$ is taken to be infinitesimal, but we cannot do that here. However because $\theta$  is constant it is possible to evaluate \eqref{FD10} exactly. Using the orthonormality condition
\begin{equation}
\int dv_x\bar{\psi}_N(x)\psi_{N^\prime}(x)=\mu^{-1}\delta_{NN^\prime},\label{FD13}
\end{equation}
it can be shown that \eqref{FD10} becomes
\begin{equation}
C_{NN^\prime}=\cos\theta\,\delta_{NN^\prime}+i\mu\sin\theta\int dv_x\bar{\psi}_N(x)\gamma_5\psi_{N^\prime}(x).\label{FD14}
\end{equation}
The integral in \eqref{FD14} is identical to the one appearing in the Fujikawa evaluation of the chiral anomaly. If we use
\begin{equation}
\ln\det\,C_{NN^\prime}={\textrm{tr}}\ln\,C_{NN^\prime},\label{FD15}
\end{equation}
by expanding the logarithm and making use of the completeness relation
\begin{equation}
\sum_{N}\psi_{N\alpha}(x)\bar{\psi}_{N\beta}(x^\prime)=\mu^{-1}\delta_{\alpha\beta}\delta(x,x^\prime),\label{FD16}
\end{equation}
where $\alpha$ and $\beta$ denote spinor indices, it can be shown that
\begin{equation}
\ln\det\,C_{NN^\prime}=i\,\mu\,\theta\,J,\label{FD17}
\end{equation}
where
\begin{equation}
J=\sum_{N}\int dv_x\bar{\psi}_N(x)\gamma_5\psi_{N}(x).\label{FD18}
\end{equation}
It is worth emphasizing that this does not assume that $\theta$ is infinitesimal. The term given by $J$ in \eqref{FD18} is exactly the same expression as occurs in the analysis of the chiral anomaly. It can be evaluated in a number of ways. If we use the heat kernel method as in \cite[Sec.~5.9]{ParkerTomsbook} then
\begin{equation}
J=i\mu^{-1}(4\pi)^{-2}\int dv_x{\textrm{tr}}\lbrack\gamma_5\,E_2(x)\rbrack,\label{FD19}
\end{equation}
where $E_2(x)$ is the heat kernel coefficient for the Dirac operator chosen in \eqref{FD11}. The necessary expression for $E_2$ was given in \eqref{3.2} and \eqref{3.3} with $W_{\mu\nu}$ and $Q$ given in \eqref{3.11} and \eqref{3.12}. Because of the presence of the $\gamma_5$ in \eqref{FD19} and the trace that occurs it can be shown using the same procedure as in \cite[Sec.~5.9]{ParkerTomsbook} that
\begin{equation}
\ln\det\,C_{NN^\prime}=\frac{i\,\theta}{768\pi^2}\,\int dv_x\,\epsilon^{\lambda\sigma\rho\tau}R_{\mu\nu\lambda\sigma}R^{\mu\nu}{}_{\rho\tau}.\label{Fd20}
\end{equation}
This result holds regardless of what is chosen for $M_0$ and $M_5$ in \eqref{3.11} and \eqref{3.12} and proves that it does not matter whether we choose the Dirac operator from \eqref{2.2b} or that from \eqref{FD7}. The net relation between $\Gamma^\prime$ and $\Gamma$ that follows from \eqref{FD12} is
\begin{equation}
\Gamma^\prime=\Gamma-\frac{1}{768\pi^2}\,\tan^{-1}\left(\frac{m_5}{m_0}\right)\int dv_x\,\epsilon^{\lambda\sigma\rho\tau}R_{\mu\nu\lambda\sigma}R^{\mu\nu}{}_{\rho\tau}.\label{FD21}
\end{equation}

The conclusion from \eqref{FD21} is that although there is an anomaly in the transformation necessary to remove the pseudoscalar mass term from the theory it is finite (no pole) and cannot affect the pole part of the one-loop effective action. This means that if the aim is to evaluate the renormalization group functions there is no loss of generality in taking only the normal scalar mass term for the spinor field. This conclusion can be verified directly from our results for the one-loop counterterms found in \eqref{5.5}. If we use the form of the action given in \eqref{FD7} we clearly get the results found from \eqref{5.5} by replacing $(m_0,m_5,w,w_5)$ with $(m_0^\prime,0,w^\prime,w_5^\prime)$. Making use of \eqref{FD6a} and \eqref{FD8} shows that the counterterms  resulting from this replacement agree precisely with those found in \eqref{5.5}.

It is also worth mentioning that instead of transforming away the pseudoscalar mass term in $m_5$ we could choose to transform away the pseudoscalar Yukawa coupling or else the scalar Yukawa coupling if we like. This follows by a suitable choice of $\theta$ in \eqref{FD1} that will involve $w$ and $w_5$ in place of $m_0$ and $m_5$. If this is done then it is necessary to consider both the scalar and pseudoscalar mass terms. The counterterms will still be invariant and there will be an anomaly of the same form as in \eqref{FD21} but where the $\tan^{-1}$ involves $w$ and $w_5$. Clearly from \eqref{FD21} there is no anomaly in flat spacetime and the presence of a pseudoscalar mass term can be ignored there. This will not be true however in the presence of a background gauge field even in flat spacetime.

\section{Conclusions}\label{sec6}

We have obtained the complete one-loop renormalization group functions for the general Yukawa model defined in Sec.~\ref{sec2}. The effective action has been obtained up to order $R^2$, including terms with two derivatives of the background scalar field, and we have discussed the possible neglect of parity violating terms in some cases. We have also analyzed how the $\bar{\psi}\gamma_5\psi$ term can be transformed away at the expense of a gravitational anomaly. (The anomaly is not present in flat spacetime which probably explains why it is not usually considered.) 

There are several extensions that could be considered. The scalar and pseudoscalar fields could be included as independent along with their associated Yukawa couplings. The fields, both scalar and spinor, could be taken to be multi-component. The local momentum space method could be used for the full effective action with general background fields and not just the interacting part as was done here. It would be possible to evaluate the renormalization group improved terms in the effective action as was done in flat spacetime \cite{ColemanandWeinberg} and for the Yukawa model in curved spacetime~\cite{elizalde1994renormalization}. The renormalization group improved effective potential in curved spacetime has been given  recently~\cite{markkanen20181}. There would be no essential impediment to proceeding to two-loop order using the local momentum space method. Finally, the inclusion of a gauge field has been performed recently~\cite{Tomsyukawa2} and the results of this will be given elsewhere.

\appendix*\section{Running value of $\lambda$}\label{append}

Here we outline the exact solution for the running value of the quartic coupling constant $\lambda$ whose renormalization group function was given in \eqref{5.2.5c}. If we define
\begin{equation}
W=w^2+w_5^2,\label{A1}
\end{equation}
then from \eqref{5.2.5i} and \eqref{5.2.5j} we have
\begin{equation}
\mu\frac{d}{d\mu}W=\frac{5}{8\pi^2}W^2.\label{A2}
\end{equation}
The solution to this is simply
\begin{equation}
W(\mu)=\frac{W_0}{1-\frac{5W_0}{8\pi^2}\ln(\mu/\mu_0)},\label{A3}
\end{equation}
where $W_0=W(\mu=\mu_0)$. This shows that $W(\mu)$ is an increasing function with a Landau pole when $\mu$ is such that the denominator is set to zero.

From \eqref{5.2.5c} we find
\begin{equation}
\mu\frac{d}{d\mu}\lambda(\mu)=\frac{3}{16\pi^2}\lambda^2(\mu)+\frac{1}{2\pi^2}W(\mu)\lambda(\mu) - \frac{3}{\pi^2}W^2(\mu).\label{A4}
\end{equation}
The solution in \eqref{A3} can be substituted for $W(\mu)$. This nonlinear equation can be turned into a linear second order differential equation by making the transformation 
\begin{equation}
\lambda(\mu)= - \frac{16\pi^2}{3}\,\mu\frac{d}{d\mu}\ln(F(\mu)).\label{A5}
\end{equation}
The equation satisfied by $F(\mu)$ is
\begin{equation}
\Big(\mu\frac{d}{d\mu}\Big)^2F(\mu)-\frac{1}{2\pi^2}W(\mu)\mu\frac{d}{d\mu}F(\mu) - \frac{9}{16\pi^4}W^2(\mu)F(\mu)=0.\label{A6}
\end{equation}
The presence of $W(\mu)$ in \eqref{A6} along with \eqref{A2} shows that there exists a solution of the form $F=W^p$ for some power $p$. If this is to solve \eqref{A6} then we require $p=p_{\pm}$ where
\begin{equation}
p_{\pm}=-\frac{1}{10}\pm \frac{\sqrt{145}}{10}.\label{A7}
\end{equation}
The solution for $\lambda(\mu)$ is easily found to be
\begin{equation}
\lambda(\mu)=-\frac{10}{3}\,W(\mu)\left\lbrack \frac{p_-(3\lambda_0+10p_+W_0)-p_+(3\lambda_0+10p_-W_0)\left(\frac{W(\mu)}{W_0}\right)^{p_+-p_-}}{3\lambda_0+10p_+W_0-(3\lambda_0+10p_-W_0)\left(\frac{W(\mu)}{W_0}\right)^{p_+-p_-}}\right\rbrack,\label{A8}
\end{equation}
where we have defined $\lambda_0=\lambda(\mu=\mu_0)$ and $W_0=W(\mu=\mu_0)$.

The asymptotic behaviour of $\lambda(\mu)$ depends on the relationship of $\lambda_0$ and $W_0$. From \eqref{A8} it can be shown that if
\begin{equation}
\lambda_0<\frac{1}{3}(1+\sqrt{145})\,W_0,\label{A9}
\end{equation}
then at some value $\mu=\mu_z$ we have $\lambda(\mu=\mu_z)=0$. The value of $\mu_z$ is given by
\begin{equation}
\ln(\mu_z/\mu_0)=\frac{8\pi^2}{5W_0}\left\lbrace 1- \left\lbrack \frac{p_-}{p_+}\left(\frac{3\lambda_0+10p_+W_0}{3\lambda_0+10p_-W_0} \right)\right\rbrack^{-\frac{1}{p_+-p_-}} \right\rbrace.\label{A10}
\end{equation}
This value of $\mu_z$ occurs before the Landau pole in \eqref{A3} is reached. It is obvious from \eqref{A4} that when $\lambda$ reaches 0 at $\mu=\mu_z$ the derivative of $\lambda$ is negative and so $\lambda$ decreases to unphysical negative values for $\mu>\mu_z$. $\lambda=0$ is not a fixed point.

In the case where \eqref{A9} is not satisfied $\lambda(\mu)$ is an increasing function of $\mu$ much as it is in the case of the real scalar field with just a quartic self-interaction.

It is possible to find the solutions for the other running coupling constants, like $\xi(\mu)$ for example, but we will not pursue this here. Details of this and application to the renormalization group improved effective potential will be given elsewhere.

\bibliography{yukawa}

\end{document}